\documentclass{pasj01} 
\usepackage{mathrsfs}
\usepackage{graphicx} 

\def\Msun{M_{\odot \hskip-5.2pt \bullet}}  \def\deg{^\circ}
  \def\hunit{{\rm H~cm}^{-2}}
\def\XCO{X_{\rm CO}} \def\Xco{X_{\rm CO}}
\def\X{X_{\rm CO:20}} 
\def\XHI{X_{\rm HI}} 
\def\ICO{I_{\rm CO}} \def\Ico{I_{\rm CO}}
\def\IHI{I_{\rm HI}}  
\def\NHI{N({\rm HI})} \def\Ioff{I_{\rm off}}
\def\xcounit{{\rm H_2~cm^{-2}(K~ km~ s^{-1})^{-1}}}
\def\Kkms{{\rm K~km~s}^{-1}} \def\kkms{{\rm K~km~s}^{-1}}
 
\def\CHISQ{$\chi^2$ } \def\CS{$\chi^2$ }  \def\tpm{ $\pm$ } 

\def\IX{I_{\rm obs}}    
\def\Ife{I_0 (E)} \def\Ibe{I_1 (E)} 
\def\If{\langle I_0 \rangle} \def\Ib{\langle I_1 \rangle} 
\def\Ioff{I_{\rm off}}
 
\begin{document}  

\title{CO-to-H$_2$ Conversion Factor of Molecular Clouds using X-Ray Shadows}  
\author{Yoshiaki \textsc{Sofue}\altaffilmark{1} and Jun \textsc{Kataoka}\altaffilmark{2}  \altaffiltext{1}{Insitute of Astronomy, The University of Tokyo, Mitaka, Tokyo 181-0015, Japan}\email{sofue@ioa.s.u-tokyo.ac.jp} \altaffiltext{2}{Research Inst. Science and Engineering, Waseda University, Shinjuku, Tokyo 169-8555, Japan} }

\maketitle 

\begin{abstract}
A new method to determine the CO-to-H$_2$ conversion factor $\XCO$ using absorption of diffuse X-ray emission by local molecular clouds was developed. It was applied to the Ophiucus (G353+17) and Corona Australis (G359-18) clouds using CO-line and soft X-ray archival data. We obtained a value $\XCO =1.85\pm 0.45 \times 10^{20} \xcounit$ as the average of least-$\chi^2$ fitting results for R4 (0.7 keV) and R5 (0.8 keV) bands.
\end{abstract}  
 
\section{Introduction}

The CO-to-H$_2$ (CO line intensity to H$_2$ column density) conversion factor, $\XCO$, is one of the fundamental parameters in the interstellar physics of molecular gas. It has been derived by several methods, which include the Virial method (Solomon et al. 1987), optical and infrared extinction method (Lombardi et al. 2008), and $\gamma$-ray method (Bloemen et al. 1986). 
 
\begin{figure} 
\begin{center}  
\includegraphics[width=6cm]{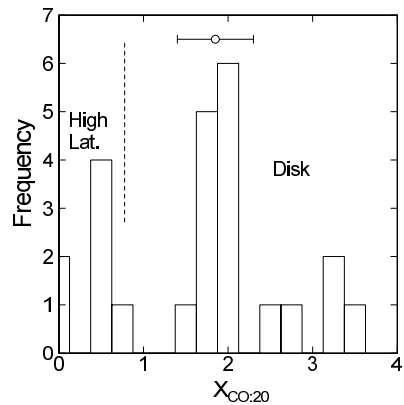}  
\end{center} 
\caption{Frequency distribution of measured $\X$ (Bolatto 2014; Moritz et al. 1998). The present resut is marked by a circle. }
\label{histo}
\end{figure}

 Although the current measurements seem to have converged to values around  $\XCO \sim 2\times 10^{20}\xcounit$, or $\X=\Xco/[10^{20}\xcounit]\sim 2$, they include still uncertainty ranging from $\X \sim 0.7$ to 4 (Bolatto et al. 2014). In fact, low values of $0.2 \sim 0.5 \times 10^{20} \xcounit$ have been obtained for the high latitude Draco cloud, cirrus, and other high latitude clouds (Moritz et al. 1998; and the literature cited therein), while a high value of $3.46\pm 0.91$ has been obtained for the Ophiucus clouds (Paradis et al. 2012). Figure \ref{histo} summarizes the $\XCO$ values obtained in the decades.

In this paper, we propose to use X-ray absorption to determine the $\XCO$ value for local molecular clouds. Low energy X-ray absorption ($\sim 0.2$ keV) has been studied for investigating the local hot plasma (Snowden et al. 2000), but few application to the conversion factor.  We developed a new method to determine $\XCO$, analyzing the CO ($J=1-0$)line data from the Colombia galactic plane CO survey (Dame et al. 2001), and soft X-ray archival data from the ROSAT observations (Snowden et al. 1997).

\section{New Method for Determining the Conversion Factor}

\subsection{Hydrogen Column Density}

We consider HI and molecular gas intervening the observer and the X-ray emission source.
The hydrogen column density $N$ along the line of sight is given by observing the integrated HI- and $^{12}$CO($J = 1-0$)-line intensities, $\IHI$ and $\ICO$, as
\begin{equation}
N = N({\rm HI})+2N({\rm H_2}),
\end{equation} 
where 
$N({\rm HI}) = \XHI \IHI$ and 
$N({\rm H_2}) = \XCO \ICO$. 
The coefficient
$\XHI = 1.82 \times 10^{18} {\rm cm^{-2}(\Kkms)^{-1}}$ is the HI intensity to H column density conversion factor, and $\XCO$ is the CO-to-H$_2$ conversion factor to be determined in this work.

\subsection{X-ray Intensity and Absorption}

 The background X-ray emission beyond a molecular cloud is assumed to be originating from the galactic halo and bulge hot plasma, and is assumed to be sufficiently uniform over the cloud extent, $\sim 10$ pc or several degrees on the sky. The foreground emission is assumed to come from the local cavity (bubble), and is also assumed to be almost flat across the clouds. 

The observed band-averaged X-ray intensity $I$ toward a cloud with column density $N$ is given by
\begin{equation}
I  =  \int \psi(E)\left[\Ife+ \Ibe e^{-N \sigma(E)}\right]dE.
\label{iobs}
\end{equation}
Here, $\Ife$ and $\Ibe$ are the foreground and background spectral intensities at photon energy $E$, $\sigma(E)$ is the absorption coefficient at $E$, and $\psi(E)$ is the response function of the instrument.  

In regions sufficiently away from the cloud, e.g. $N\simeq 0$, the off-cloud band-averaged (observed) intensity is written as
\begin{equation}
\Ioff=\If+\Ib,
\end{equation}
 where
\begin{equation}
\If=\int\psi(E)\Ife dE
\end{equation}
and
\begin{equation}
\Ib=\int\psi(E)\Ibe dE.
\end{equation}   

For a later convenience we rewrite equation (\ref{iobs})
as
\begin{equation}
I =\If  + \Ib \int \psi(E) \eta(E) e^{-N \sigma(E)} dE,
\label{iobsb}
\end{equation} 
where 
\begin{equation}
\eta(E)=\Ibe/\Ib
\end{equation}
represents the normalized spectrum of the background emission,
satisfying
\begin{equation}
\int \psi(E)\eta(E)dE = 1.
\label{eta}
\end{equation}
The function $\eta(E)$ is calculated for an arbitrary emission measure
assuming the temperature and metallicity using XSPEC
(Arnauld et al. 1996) as shown in figure \ref{spec}.
In the current shadow analysis, the so-called 'band-averaged cross section' is used (Snowden et al. 2000)
\footnote{
The 'band-averaged cross section' is defined through
$ I=\If+ \Ib e^{-N\sigma'},  $ and is expressed as
$   \sigma'  =(1/ N){\rm ln} [ {\int \psi(E) I_1(E) e^{-N\sigma(E)} dE}/ \Ib ].
$ 
}, which is shown to be equivalent to the present direct band-averaging method.  

The instrument response was taken from Snowden et al. (1997), and was approximated by Gaussian functions around the band centers at $E_0=0.70$, 0.83 and 1.11 keV with full widths of half maximum of $\Delta E=0.39$, 0.45, and 0.40 keV, respectively. This approximation was good enough in the present accuracy of analysis.  

\subsection{Emission and Absorption Spectra}

\begin{figure} 
\begin{center} 
\includegraphics[width=7cm]{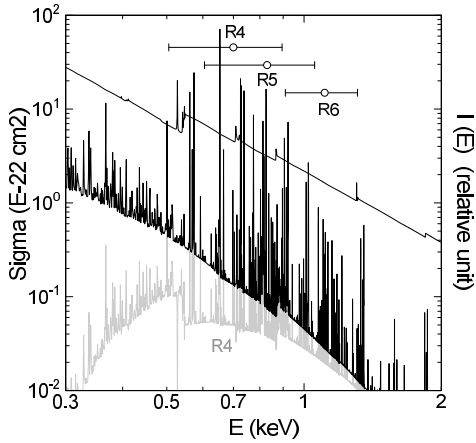}   
\end{center} 
\caption{
[Upper lines] Absorption cross section $\sigma(E)$ for $Z=Z_\odot$ created by 'ISMabs' (Gatuzz et al. 2015) implemented in XSPEC (Arnauld et al. 1996). ROSAT bands are indicated by the circles.
[Middle line] X-ray spectrum at $kT=0.3$ keV for $Z=0.2 Z_\odot$ with abundance ratios by Anders and Grevese (1989) for arbitrary $EM$ using XSPEC. 
[Lower gray line] Same but multiplied　by exp$(-\sigma/\sigma_4)$, showing that transmitted spectrum is flat near $\tau \sim 1$ in R4 band. }
\label{cs}
\label{spec}
\end{figure}

Figure \ref{cs} shows $\sigma(E)$ plotted against $E$ for neutral gas with solar abundance, as generated by using 'ISMabs' (Gatuzz et al. 2015) implemented in XSPEC (Arnauld et al. 1996). We also show a cross section for ionization degree as high as 10\%, unrealistically high in interstellar clouds, in order to confirm that the contribution of ions is negligible. 

The figure also shows an emission spectrum created using XSPEC for an arbitrary fixed emission measure of plasma with temperature $kT=0.3$ keV and metallicity $Z=0.2Z_\odot$, representing the galactic halo emission (Kataoka et al. 2013, 2015). The spectral shape was used in the fitting as the weighting function, and the intensity scale is so normalized that the band-averaged intensity represents the observed intensity by equation (\ref{iobs}). 
 
The gray line in figure \ref{cs} shows a spectrum of transmitted emission in the R4 band through a cloud with optical depth unity ($\tau=1$), as obtained by multiplying exp$(-\sigma/\sigma_4)$ to the emissin spectrum in order to demonstrate that the transmitted spectrum is flat in regions with $\tau\sim 1$, where the absorption analysis is most efficiently obtained.


We comment on the assumed metallicities:  (a) The foreground emissin is subtracted before fitting, and hence its metallicity is not included. (b) Metallicity variation in the background halo and bulge is assumed to be small across the cloud extents. (c) Metallicity of molecular clouds is not known, but is usually approximated by that of nearby stars or HII regions. As the clouds are near the Sun, we assume the solar metallicity.

 \subsection{$\chi^2$ Fitting}
 
Equations (\ref{iobs}) and (\ref{iobsb}) can be solved for the two unknown parameters, $\XCO$ and $\If$ (or $\If=\Ioff-\Ib$), by measuring intensities $\IX$ at multiple positions. Practically, we determine the parameters that minimize \CHISQ defined by 
\begin{equation}
\chi^2 =\Sigma \left[\IX - I \right]_i^2/s_i^2,
\label{eqchi}
\end{equation}
where $\IX$ and $s_i$ are observed X-ray intensity and dispersion at the $i$-th measured position, respectively, and $I$ is calculated using equation (\ref{iobs}) for observed $\Ico$ at the same position corresponding to the assumed parameter pair of $\XCO$ and $\If$.

\begin{figure} 
\begin{center} 
\includegraphics[width=7cm]{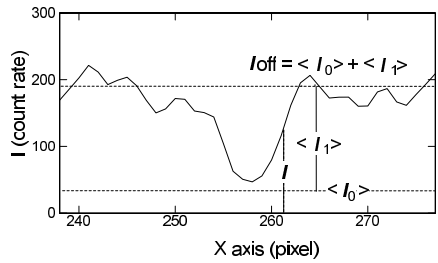}  
\end{center} 
\caption{Intensity variation across the molecular cloud CrA in R4 band and the definition of variables.}
\label{cloudcut}
\end{figure}

The off-cloud intensity $\Ioff=\Ib+\If$, representing the baseline value, was measured as the mean of $I$ at positions sufficiently away from the cloud. Figure \ref{cloudcut} illustrates the definition of the variables using an intensity variation across Cr A molecular cloud in R4 band. More practically, we measured an average of intensity $I$ at positions where $\ICO $ is sufficiently small with $\ICO \le I_{\rm CO:off}= 0.5 \kkms$. 

\section{Molecular Clouds} 

\begin{figure*} 
\begin{center}
\includegraphics[height=7cm]{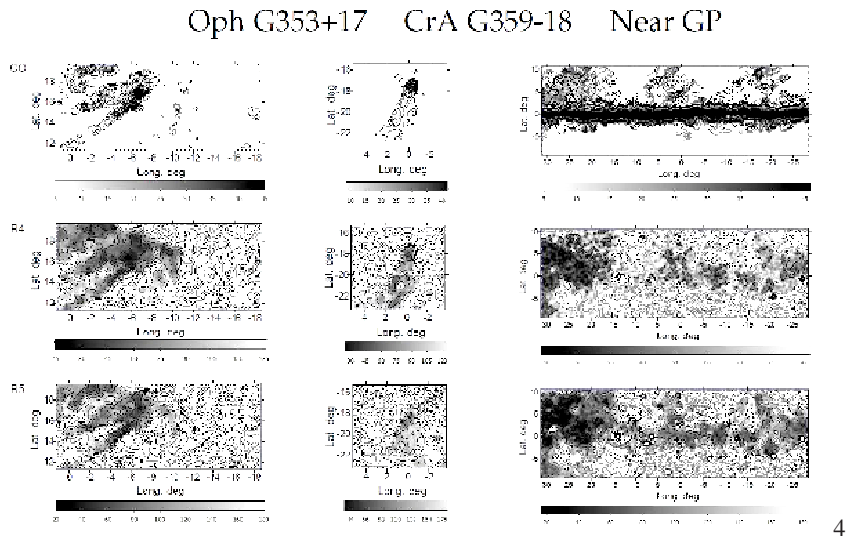} 
\ref{clouds}
\end{center} 
\caption{CO (from Columbia survey), X-ray R4 and R5 band (from ROSAT survey) intensity maps of Oph (G353+17), CrA (G359-18) molecular clouds, and near galactic plane region, where only points with $\Ico \le 20 ~\Kkms $ were used in order to avoid the galactic plane at $|b|<\sim 3\deg$. 
}
\label{clouds}
\end{figure*}

Figure \ref{clouds} shows intensity distributions made from Columbia CO line survey and ROSAT X-ray maps in R4 (0.7 keV), 5 (0.8) and 6 (1.1) bands for a molecular cloud G353+17 (Ophiucus), comet-like cloud G359-18 (CrA), and a near-galactic plane region. The R2 band was not used for too heavy absorption, nor R6 and 7  for noisier data particularly around CrA. 
The figures exhibit beautiful anticorrelation between the molecular gas and X-ray intensities, demonstrating that the global X-ray structures are determined by the interstellar absorption. This fact implies that the intrinsic X-ray distribution is sufficiently flat compared to the intensity fluctuation due to cloud extinction. 

\subsection{Ophiucus molecular cloud G353+17}

The Ophiucus molecular cloud G353+17 (de Geus and Burton 1991) was chosen, because the surrounding X-rays are bright and diffuse enough. The cloud is isolated, and its distance is known to be $d = 119\pm 6$ pc (Lombardi et al. 2008). The densest region is located at G353.0+16.7 (Liseau et al. 2010).
Figure \ref{clouds} shows that the shape of
the soft X-ray shadow almost exactly coincides with that of the molecular gas distribution. 

The space between the cloud and the Sun is almost empty, composing the local cavity (Puspitarini et al. 2014). The dust extinction in the cavity is as low as  $E(B-V ){\rm pc}^{-1} \sim 10^{-4} {\rm mag~ pc}^{-1}$ corresponding to $N({\rm H}) \simeq 5\times 10^{21} E(B-V)d \simeq 0.6\times 10^{20} \hunit$. Thus, we assume that HI inside the cavity is constant at $N(HI)=0.6 \times 10^{20}\hunit$.

 Beyond and surrounding the Oph cloud, HI is distributed smoothly compared to the cloud size  at intensity $\IHI\sim 300 ~\kkms$ or $ \NHI \sim 5\times 10^{20} {\rm cm}^{-1}(\kkms)^{-1}$ (Kalberla et al. 2005). This diffuse HI gas may suppresses the intrinsic X-ray background $I_1$ by $\sim 10$\%.

The HI contribution within the cloud is negligible, as the H$_2$ to HI column density ratio is observed to be N(H2)/N(HI)=56 in the Oph cloud by HI self absorption (Minn et al. 1981), or HI contributes only $\sim 2$\%, which, however, does not affect the fitting after the on-off position subtraction.

\subsection{Corona Australis cloud G359-18}

G359-18 cloud in Corona Australis (CrA) is a well known southern comet-like dark cloud associated with the reflection nebula NGC 6729 and B stars as well as infrared emission (Odenwald 1988). It is located at a distance of $129 \pm 11$ pc (Casey et al. 1998).  The cloud appears as an isolated shadow against nearly uniform bulge X-ray emission. We here assume the same HI condition as for the Oph cloud.

\subsection{Near galactic plane region}

We also applied the method to a near-plane region as shown in figure \ref{clouds}. The gas distribution is well anti-correlated with the X-ray emission. However, we here consider the result only for reference, as the area is too wide for the assumption of a flat X-ray background. In the analysis we used only data points with $\Ico \le 20~\Kkms$ in order to avoid the galactic plane region at $|b|<\sim 3\deg$ (dark CO regions in figure \ref{clouds}).

\section{Least $\chi^2$ Fitting for $\XCO$}

\subsection{TT Plot}

\begin{figure} 
\begin{center}  
Oph\\
\includegraphics[width=8cm]{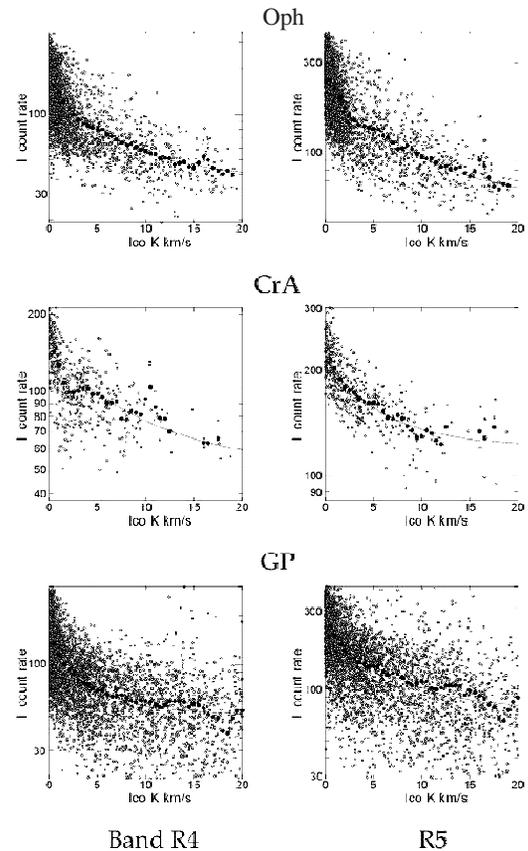}  
\end{center} 
\caption{ X-ray intensities plotted against CO intensities  (TT plot).}
\label{TT}
\end{figure}

Figure \ref{TT} shows plots of the observed X-ray intensity ($\IX$) against CO-line intensity ($\ICO$). Gray dots are raw values at individual pixels, and full circles are Gaussian-averaged values in $\ICO$ bins each 0.5 K km s$^{-1}$ used as observed $I_{\rm obs}$ corresponding to $\ICO$ in equation (\ref{eqchi}). The plotted $\IX$ values at $\ICO \le 20\kkms$ can be well approximated by a smooth curve representing the absorption. The $\chi^2$ fitting is thus applied to data with $\ICO \le 20 ~\kkms$. 
 
\subsection{\CS Fitting}

In order to apply the $\chi^2$ fitting, we first measured the off cloud X-ray intensities, $\Ioff$, as the mean of observed values at positions where the CO intensity is less than 0.5 $\kkms$. Using equation (\ref{eqchi}), we apply the \CS fitting to the averaged intensities $\IX$ and $\ICO$ indicated in figure \ref{TT} and associated standard deviations $s_i$. Figure \ref{chi} shows the variation of calculated \CS value as a function of $\Xco$ for different $\Ib$ values near the \CS minimum.

\def\h{\hskip -5mm}
\begin{figure} 
\begin{center} 
Oph\\
\includegraphics[height=2.5cm]{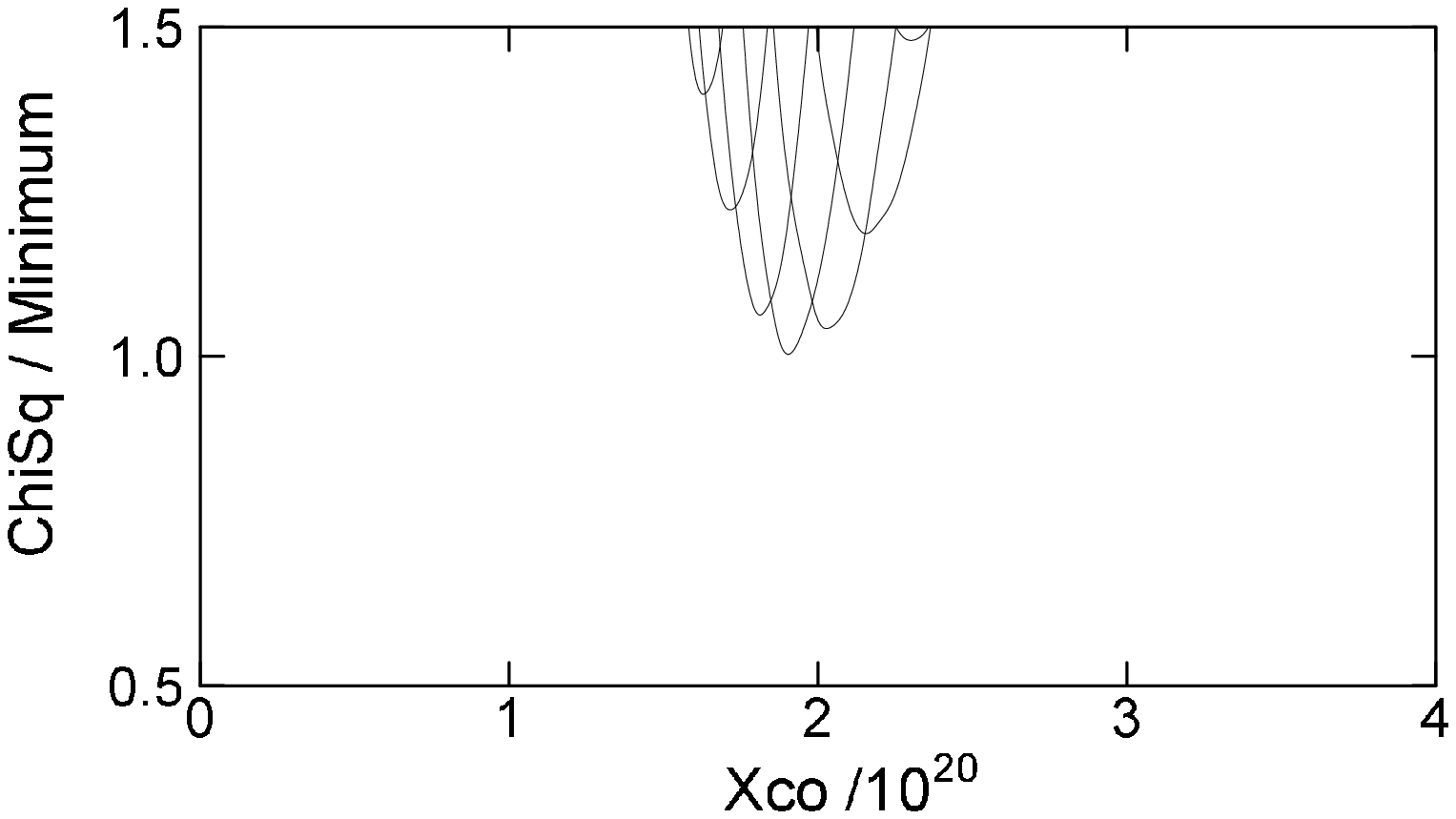}
\h \includegraphics[height=2.5cm]{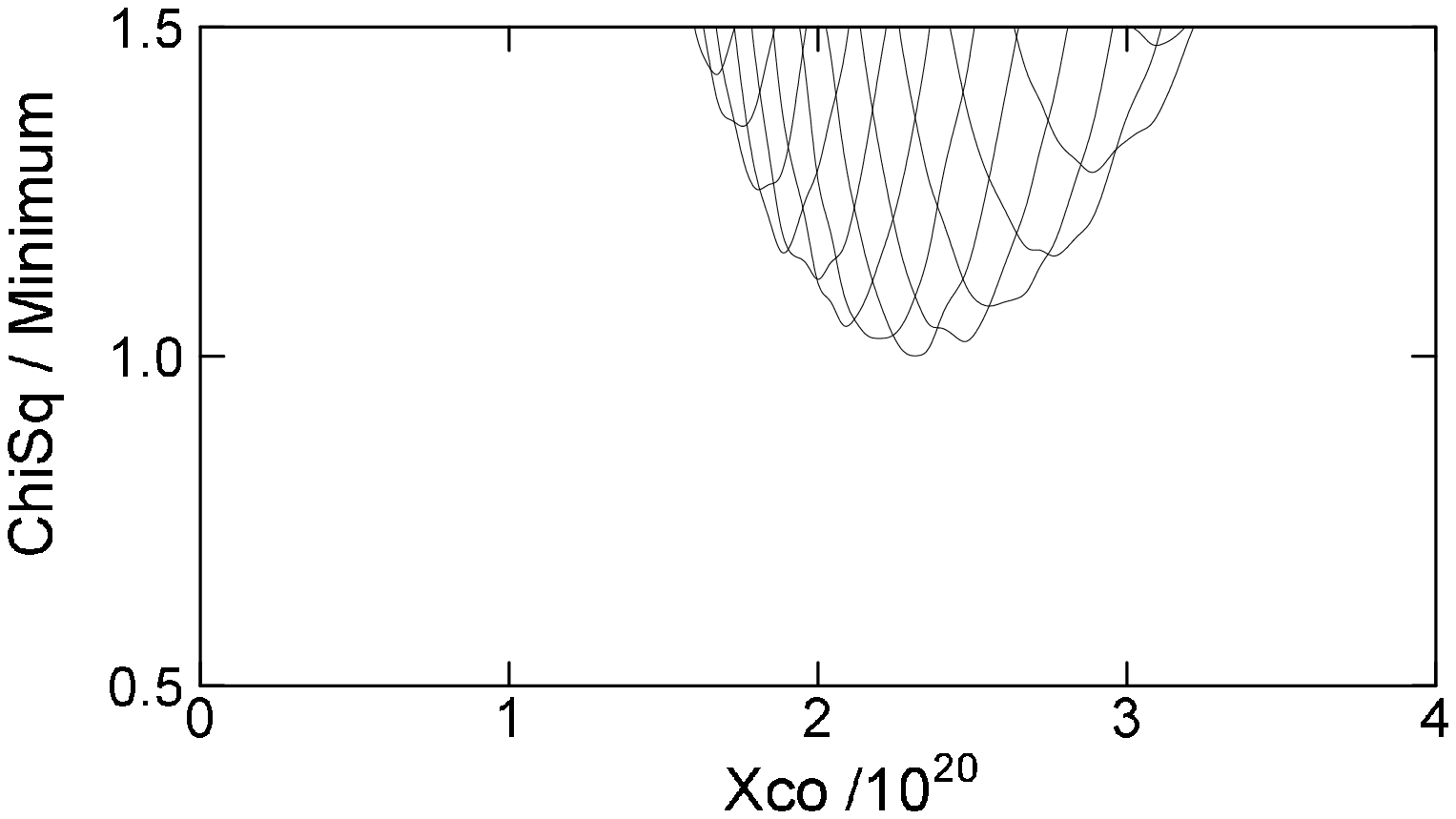}\\
CrA\\
\includegraphics[height=2.5cm]{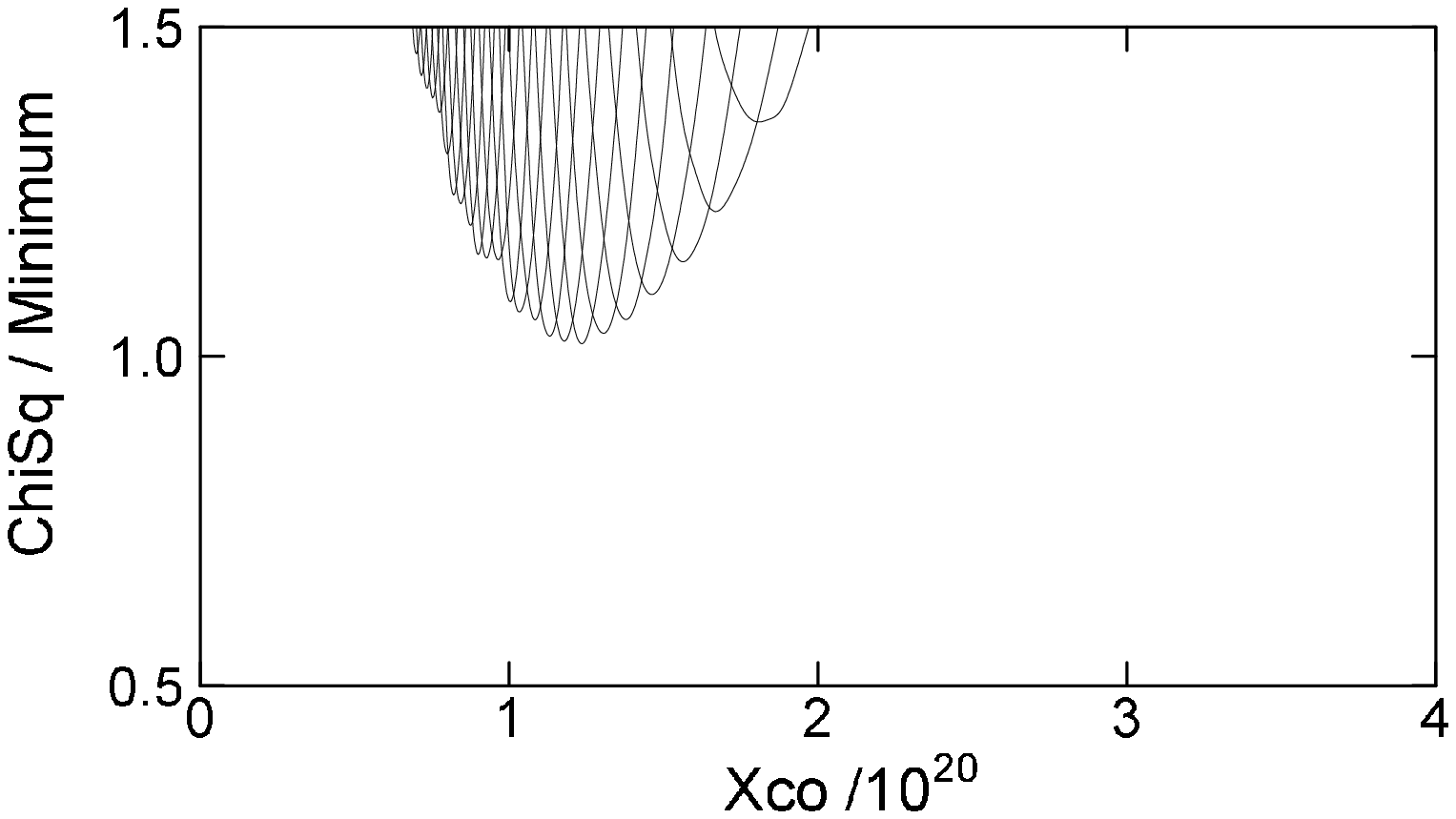}
\h \includegraphics[height=2.5cm]{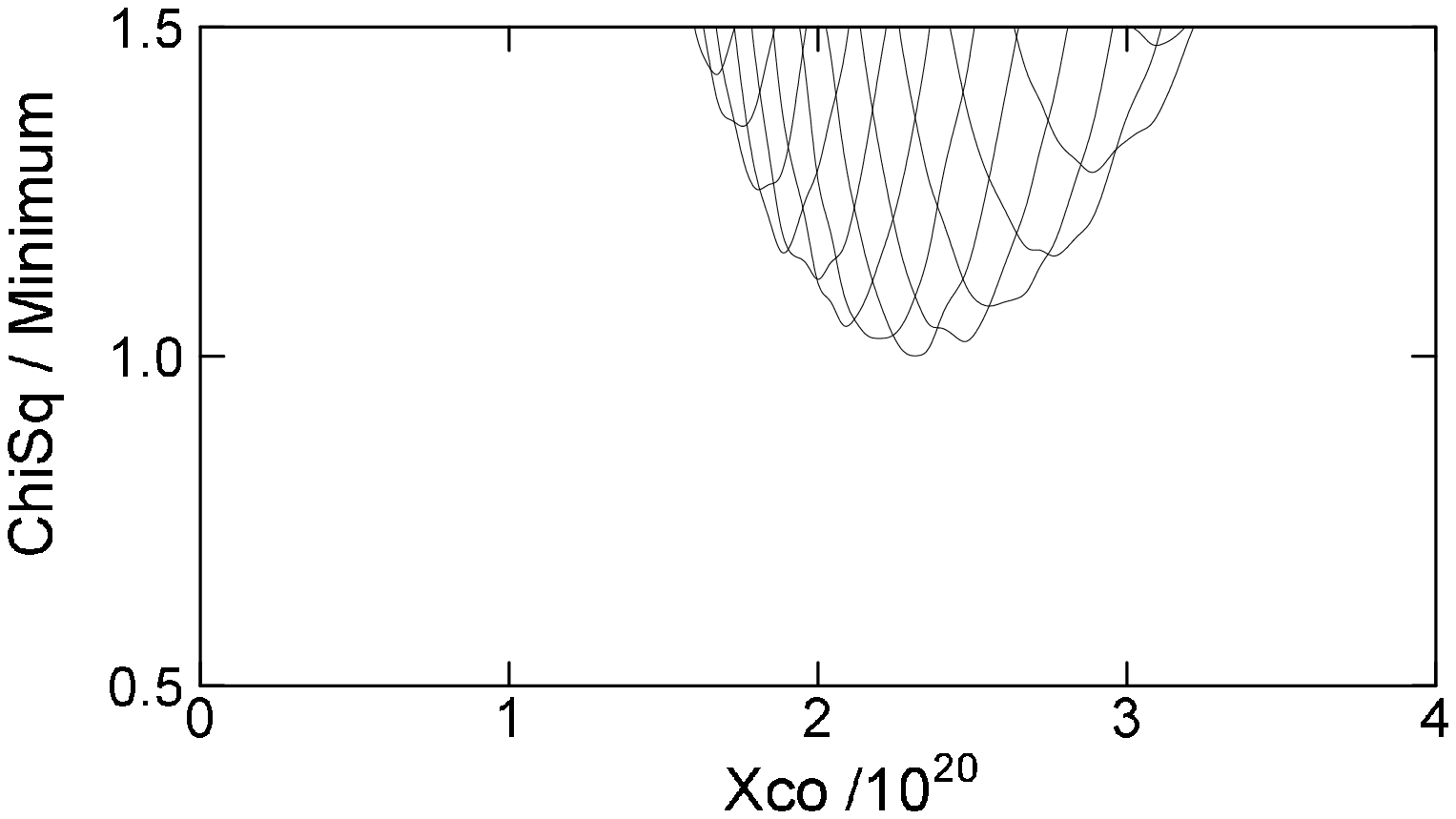}\\
GP\\
\includegraphics[height=2.5cm]{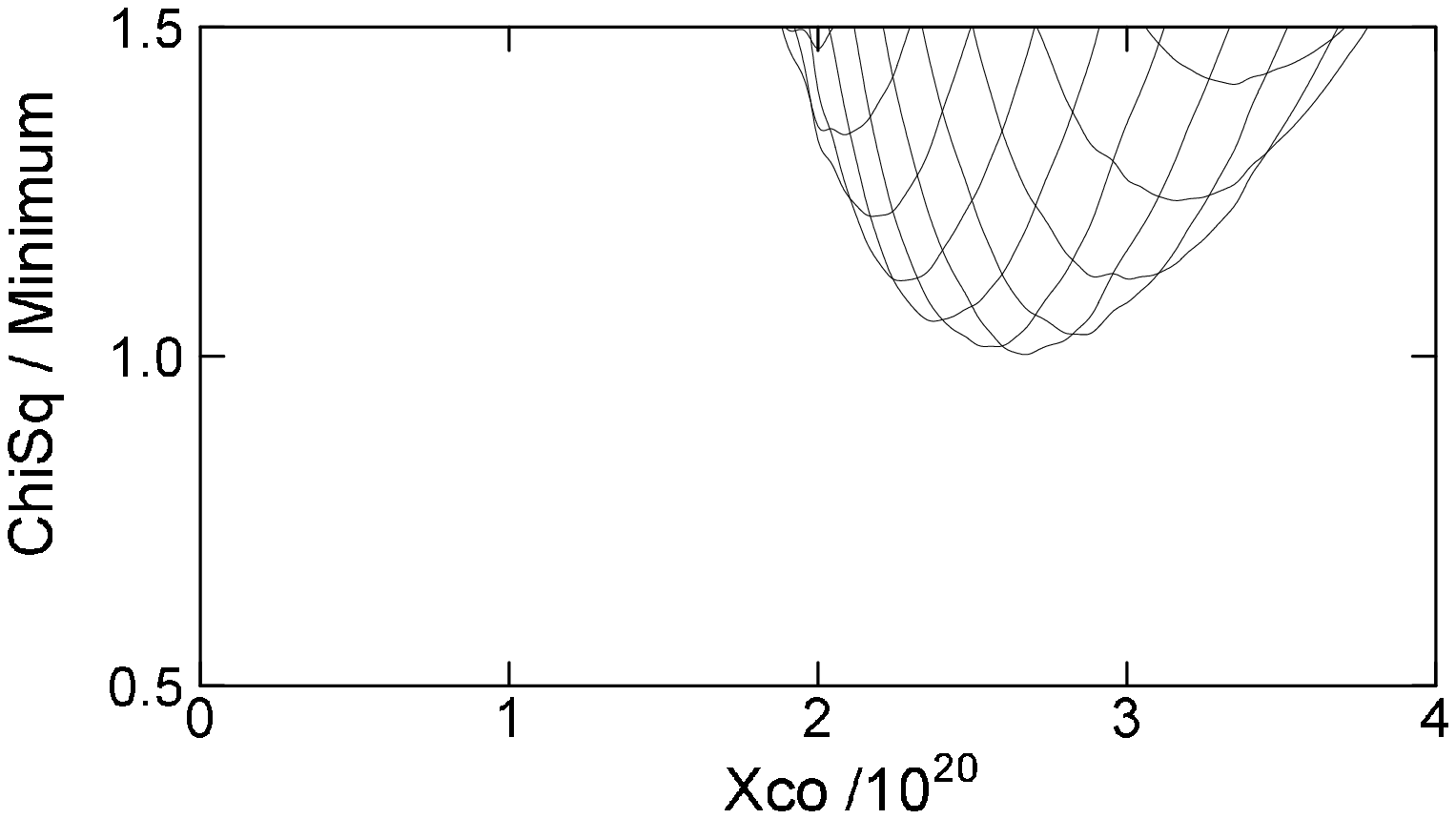}
\h \includegraphics[height=2.5cm]{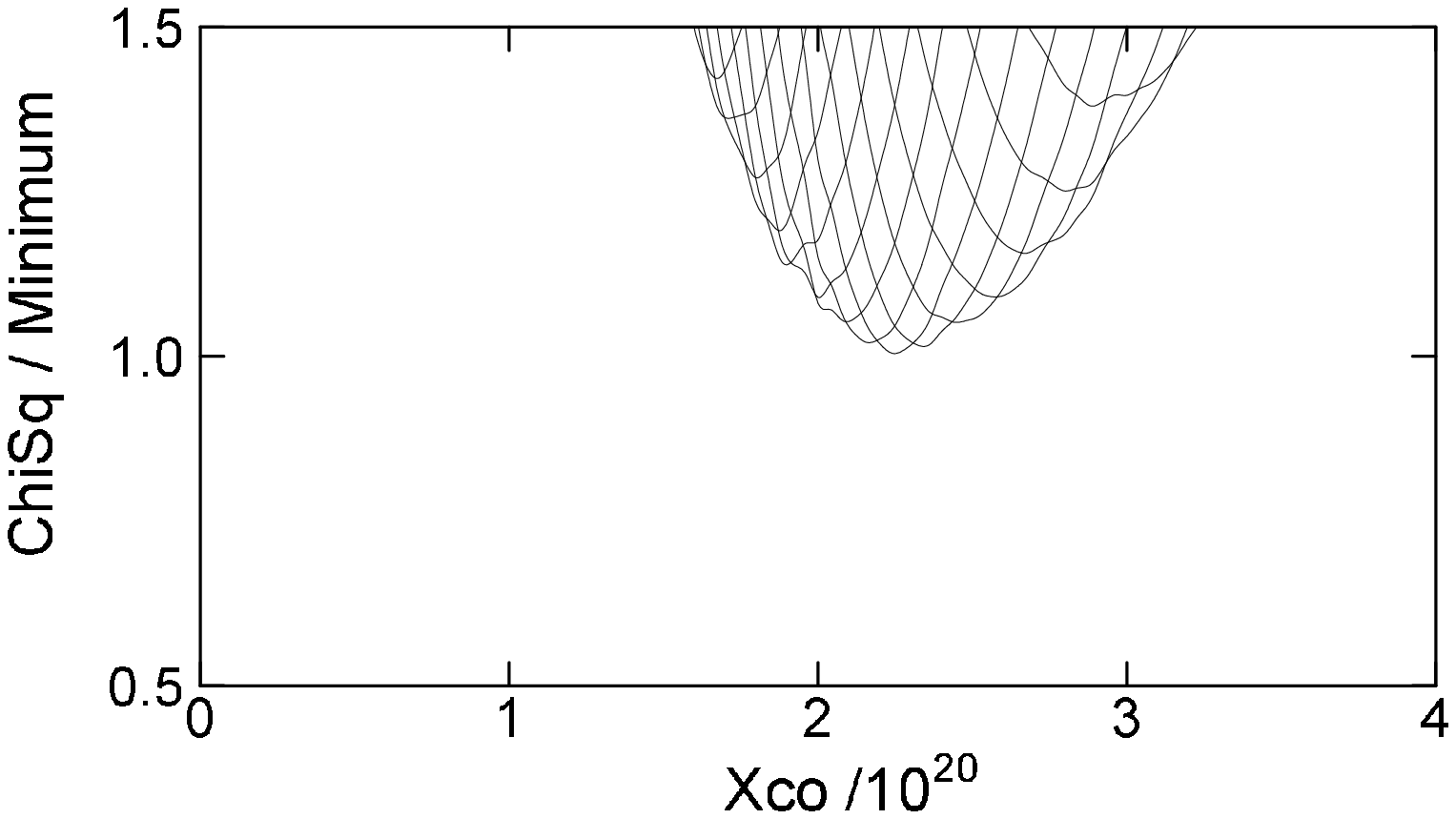}\\
Band R4 ~~~~~~~~~~~~~~~~~~~~~~~~~~~~~~~~~~ R5 
\\ 
\end{center} 
\caption{Variation of \CS with $\XCO$ in unit of $10^{20} \xcounit$ around the minimum for Oph and CrA clouds. 
 Curves correspond to different $I_1$ around the minimum.}
\label{chi}
\end{figure} 

The \CS~ values were computed for various combinations of the two parameters $\Ib$ and $\XCO$, and are shown in figure \ref{chi}. The best fitting value for $\XCO$ was found by fitting a parabola to the three pairs of $(\Ib, \XCO)$ in the region of minimum \CS. The fitting error was calculated as the range of $\XCO$ that allows an increase of \CS\ by 1 for a fixed $\Ib$ near the \CS\ minimum. 

Table \ref{tabres} lists the \CS fitting results. The results for R6 band are largely scattered due to the noisier data compared to the other two bands. Thus, we adopted the R4 and R5 band values, and averaged them to yield a mean value of $\X=1.85\pm 0.45$.  

\begin{table}
\caption{CO-to-H$_2$ conversion factor $\XCO$ by least-\CS fitting of X-ray absorption method.}
\begin{center}
\begin{tabular}{lll}  
\hline \hline
Clouds& Band &$\X$ \\
\hline 
Oph G353+17 
&R4 &  1.90 \tpm    0.16$^\dagger$ \\
&R5 &1.96 \tpm    0.18   \\
\hline
CrA G359-18 
&R4& 1.24 \tpm    0.06 \\
&R5 &  2.31 \tpm    0.57  \\
\hline \hline
Average& R4+5 & $1.85\pm 0.45^\ddagger$ \\  
\\
\hline \hline
Near GP for reference\\
\hline
 &R4& 2.67 \tpm    0.41  \\
&R5 &  2.24 \tpm    0.36  \\
\hline   
\end{tabular}
\end{center}
$^\dagger$ Uncertainty allowing for \CS$\le$\CS(min)+1 estimated by parabolic fitting of the \CS~ function near the minimum.\\
$^\ddagger$ Standard error of average of the above values. 
\label{tabres} 
\end{table}

\section{Discussion}

We described the \CS ~fitting method to determine the CO-to-H$_2$ conversion factor using X-ray shadows by molecular clouds. The method was applied to the Ophiucus and CrA molecular clouds, which were so chosen that they are conveniently located in the direction of the diffuse X-ray halo and bulge whose intrinsic distribution is considered to be sufficiently uniform within the analyzed scales. The two clouds are local and hence, the $\XCO$ values apply to clouds with the solar metallicity.

By minimizing \CS, the CO-to-H$_2$ conversion factor was determined as listed in table \ref{tabres}, and the conversion factor determined to be $\X=1.85\pm 0.45$ as the average value for the two clouds. The value is compared with the current values in figure \ref{histo}, and is close the mean of the disk values. This is consistent with the fact that the two clouds are local at heights $z\sim \pm 30$ pc from the galactic plane. 

 We also tried to apply the method to a near-galactic plane region, but only for reference, as the assumption of uniform X-ray background may not hold in this large area. Nevertheless, we obtained consistent $\Xco$ values within the larger errors as in table \ref{tabres}. 

 Figure \ref{histo} shows a frequency distribution of measured $\XCO$ values, mostly from the well summarized table by Bolatto et al. (2013), where low values $\X<0.5$ in high latitude clouds  (Moritz et al. 1998) are also plotted . The figure exhibits two groups of $\X$ values around $\sim 2$ and  $<\sim 0.5$. This might suggest that the disk and high-latitude clouds have different chemical properties. 
 
  The present X-ray shadow method may be an independent tool to measure $\Xco$ in individual molecular clouds. It will be worth to compare our results with other measurements.
 \begin{itemize}
 \item Oph cloud:  de Geus and Burton (1990) obtained a Virial mass of the Oph cloud as large as $5-7 \times 10^4 \Msun$, an order of magnitude more massive than the converted mass of $9 \times 10^3\Msun$ for an assumed $\X=2.6$. From this discrepancy, they argue that the Oph cloud will not be virialized. This may imply that the present shadow method can be a useful tool to measure $\Xco$ in non-virialized clouds. 
 \item CrA cloud: Ackermann et al. (2012) obtained $\X=0.99 \pm 0.08$ by applying the $\gamma$-ray method to their Fermi data. This value is consistent with our R4 band value of $1.24\pm 0.06$, and is not inconsistent with the R5 value of $2.31\pm 0.57$.
\item
Near Galactic plane: Ade et al. (2011) applied the dust-extinction method to the Planck data for galactic dark clouds, and obtained $\X=2.54 \pm 0.13$. Paradis et al. (2012) obtained a relatively high value, $3.46 \pm 0.91$, for Oph-Aauila region at $b\ge 10\deg$ using the dust extinction method. These regions are partly overlapping with our GP area, and their $\X$ values are consistent with our R4 and R5 values, $\X = 2.67\pm 0.41$ and $2.24\pm 0.36$, respectively. 
\end{itemize}

Finally we comment on the limitation of the present method. Constant foreground HI emission, $0.6\times 10^{20}\hunit$, was assumed in the local cavity. However, an increase of this value by $0.1\times 10^{20}$ will decrease $\XCO$ by $\sim 1$\%, and vice versa.  
 Absorption by cold HI gas within molecular clouds was shown to be negligible in the  Oph cloud, which may also apply to CrA. On the other hand, contribution of cold HI may be more serious when the method is applied to dense and more massive clouds as GMCs. 
Another uncertainty arises from the intrinsic inhomogeneity of the back- and fore-ground X-ray emissions. This difficulty could be eased by applying the method to smaller clouds at higher angular resolutions, providing with more accurate determinations.

\vskip 5mm \noindent{\it Acknowledgements}: The analysis made use of archival data of the ROSAT all-sky survey and the Columbia galactic CO line survey. We would like to thank Drs. E. Gatuzz and J. Garcia for useful advice on ISMabs code implemented in the XSPEC. We also thank Mr. M. Akita and T. Mimura  for  their technical support in the spectral simulation.

\end{document}